\documentclass[12pt]{iopart}
\usepackage{etoolbox}

\makeatletter
\renewcommand\@appendixstar{\@@par
 \ifnumbysec 
 \@addtoreset{table}{section}
 \@addtoreset{figure}{section}\fi
 \setcounter{section}{0}
 \setcounter{subsection}{0}
 \setcounter{subsubsection}{0}
 \setcounter{equation}{0}
 \setcounter{figure}{0}
 \setcounter{table}{0}
 \def\thesection{\Alph{section}} % this line has been \def\thesection{Appendix \Alph{section}} before
 \def\theequation{\ifnumbysec
      \Alph{section}.\arabic{equation}\else
      \Alph{section}\arabic{equation}\fi}
 \def\thetable{\ifnumbysec
      \Alph{section}\arabic{table}\else
      A\arabic{table}\fi}
 \def\thefigure{\ifnumbysec
      \Alph{section}\arabic{figure}\else
      A\arabic{figure}\fi}}
\makeatother

\expandafter\let\csname equation*\endcsname\relax
\expandafter\let\csname endequation*\endcsname\relax

\usepackage[fleqn]{amsmath}
\usepackage[]{amssymb}
\usepackage[]{mathrsfs}
\usepackage[]{tensor}
\usepackage[]{amsthm}
\usepackage[]{bm}
\usepackage[normalem]{ulem}

\usepackage{pgfplots,epsfig}
\pgfplotsset{/pgf/number format/use comma,compat=newest}

\numberwithin{equation}{section}

\newcommand{\0}{\nonumber}
\newcommand{\dd}{\partial}
\newcommand{\df}{\mathrm{d}}
\newcommand{\defeq}{\mathrel{\mathop:}=}

\newcommand{\Hodge}{\star}
\newcommand{\Lie}{\pounds}
\newcommand{\nab}[1]{\nabla_{\!#1}}

\DeclareMathOperator\const{const}

\def\bc#1\ec{\begin{center}#1\end{center}}
\def\be#1\ee{\begin{align}#1\end{align}}

\usepackage[unicode]{hyperref}
\usepackage{xcolor}
\definecolor{pastgreen}{HTML}{669900}
\definecolor{pastblue}{HTML}{336699}
\definecolor{pastred}{HTML}{990000}
\definecolor{linkcol}{HTML}{663333}
\hypersetup{colorlinks,linkcolor={pastblue},citecolor={pastblue},urlcolor={pastblue}}

\theoremstyle{plain} \newtheorem{tm}{Theorem}[section]
\theoremstyle{plain} \newtheorem{lm}[tm]{Lemma}
\theoremstyle{definition} \newtheorem{defn}[tm]{Definition}
\theoremstyle{definition} 

\def\btm#1\etm{\begin{tm}#1\end{tm}}
\def\blm#1\elm{\begin{lm}#1\end{lm}}
\def\bdefn#1\edefn{\begin{defn}#1\end{defn}}

\usepackage[]{cite}

\usepackage[capitalize]{cleveref}

\makeatletter\g@addto@macro\bfseries{\boldmath}\makeatother%

\def\keywords#1{\vspace{10pt}
\begin{indented}
\item[]\rm Keywords: #1\par
\end{indented}}

\begin{document}

\begin{flushright}
ZTF-EP-21-01
\end{flushright}

\title{Stationary spacetimes with time-dependent real scalar fields}

\author{Edgardo Franzin$^{a,b,c,d}$ and Ivica Smoli\'c$^e$}
\address{$^a$ Department of Astrophysics, Cosmology and Fundamental Interactions (COSMO), Centro Brasileiro de Pesquisas F{\'\i}sicas (CBPF), rua Dr.\ Xavier Sigaud 150, Urca, Rio de Janeiro -- RJ, 22290-180 Brazil}
\address{$^b$ SISSA, International School for Advanced Studies, via Bonomea 265, 34136 Trieste, Italy}
\address{$^c$ INFN, Sezione di Trieste, via Valerio 2, 34127 Trieste, Italy}
\address{$^d$ IFPU, Institute for Fundamental Physics of the Universe, via Beirut 2, 34014 Trieste, Italy}
\address{$^e$ Department of Physics, Faculty of Science, University of Zagreb, Bijeni\v cka cesta 32, 10000 Zagreb, Croatia}
\eads{\mailto{efranzin@sissa.it}, \mailto{ismolic@phy.hr}}

\date{\today}

\begin{abstract}
In 1981 Wyman classified the solutions of the Einstein--Klein--Gordon equations with static spherically symmetric spacetime metric and vanishing scalar potential. For one of these classes, the scalar field linearly grows with time. We generalize this symmetry noninheriting solution, perturbatively, to a rotating one and extend the static solution exactly to arbitrary spacetime dimensions. Furthermore, we investigate the existence of nonminimally coupled, time-dependent real scalar fields on top of static black holes, and prove a no-hair theorem for stealth scalar fields on the Schwarzschild background.
\end{abstract}



%%%%%%%%%%%%%%%%%%%%%%%%%%%%%%
%%%%%%%%%%%%%%%%%%%%%%%%%%%%%%
\section{Introduction} %%%%%%%
%%%%%%%%%%%%%%%%%%%%%%%%%%%%%%
%%%%%%%%%%%%%%%%%%%%%%%%%%%%%%

Most of the exact gravitational solutions are derived using Ans\"atze constructed with symmetries but, generically, solutions of gravitational field equations ought not to possess any symmetries, that is, to admit any Killing vector fields.
When they do, it is usually assumed that all nongravitational fields present in the solution \emph{inherit} the spacetime isometries:
For any Killing vector field $K^a$, such that $\Lie_K g_{ab} = 0$, we say that a physical (tensor) field $\tensor{\Psi}{^{a\dots}_{b\dots}}$ inherits the corresponding symmetry if $\Lie_K \tensor{\Psi}{^{a\dots}_{b\dots}} = 0$.
This assumption, however, is not necessary and its circumvention may lead to physically interesting solutions.

The investigation of symmetry inheritance and symmetry noninheriting field configurations has a long history, with an early burst of activity going back to the 1970s~\cite{Woo73a,Woo73b,MzHRS74,MW75,Coll75,RT75,WY76a,WY76b}.
These papers were mostly focused on four-dimensional electromagnetic fields~$F_{ab}$ and the main conclusion was that $\Lie_K F_{ab} = f\,{\Hodge F_{ab}}$, where the real function~$f$ is indeed nonvanishing in some electrovacuum spacetimes~\cite{MW75,WY76b,LP77,FC78}.
Subsequent generalizations include e.g.\ investigations of the additional constraints in presence of black-hole horizons~\cite{Tod06}, proofs that lower-dimensional electromagnetic fields necessarily inherit the symmetries~\cite{CDPS16,BGS17} and some initial results on nonlinear electromagnetic models~\cite{BGS17}.

The research on symmetry inheritance for scalar fields had a more modest start, with the pioneer works by Hoenselaers~\cite{Hoen78} and Wyman~\cite{Wyman81}, followed by a long dormant period.
Investigating the Einstein--Klein--Gordon equations with a static, spherically symmetric spacetime metric, Wyman presented two classes of solutions. In the first and best known class of solutions, discovered originally by Fisher~\cite{Fisher48}, the scalar field inherits all spacetime isometries.\footnote{The first class of solutions serves to illustrate the so-called Stigler's law of eponymy, as it has been rediscovered many times over, first by Fisher, then Bergmann and Leipnik~\cite{BL57}, Janis, Newman and Winicour~\cite{JNW68}, and Buchdahl~\cite{Buchdahl72}, before Wyman, and discovered again by Agnese and La Camera~\cite{Agnese:1982,Agnese:1985xj}, Dionysiou~\cite{Dionysiou:1982} and Roberts~\cite{Roberts:1993re}.}
In the second class, the scalar field does not inherit the spacetime isometries and it is a linear function of time.
Wyman explicitly gave one global and one power-series approximate solutions to which we refer, respectively, as Wyman~IIa and IIb.
In the cosmological extension of the Wyman~IIb solution found by Pi~\cite{Pi05} and Jackiw~\cite{Jackiw06}, the scalar field is no longer symmetry noninheriting, as the metric itself is not stationary.
Nonetheless, there exist static cosmological extensions of the Wyman~IIa solution with a scalar field which linearly grows with time~\cite{Maeda12,KG15,Sultana15}, in accordance with the general rules of symmetry inheritance~\cite{ISm15,ISm17}.
We note in passing that the Wyman~II solutions admit a simple ``complexification''~\cite{ISm15}, and that the cosmological extension of the Wyman IIa solution also has a Brans--Dicke counterpart~\cite{BFF19}.

A more recent wave of activity was initiated by the discovery of nontrivial complex scalar field configurations which can coexist with rotating black holes~\cite{HR14,HR15}.
These solutions evade classical no-hair theorems precisely owing to symmetry noninheritance of the scalar field. A thorough analysis of the symmetry inheritance properties of minimally coupled real and complex scalar fields~\cite{ISm15} has shown off strict constraints (see also \cite{Graham:2014ina}), especially enhanced in presence of Killing horizons~\cite{ISm17}.
In other words, we cannot evade earlier no-real-scalar-hair theorems via symmetry noninheritance, at least as long as the scalar field is minimally coupled.
The first attempt to generalize some of these results to nonminimally coupled scalar fields~\cite{BS17} offered an insight into a classification scheme which is still incomplete. One of the most important open questions is about the possible forms of black-hole solutions with a symmetry noninheriting real scalar hair nonminimally coupled to gravity.

Hence, the scope of this paper is manifold, closing remaining gaps in the theory of symmetry noninheriting minimally coupled real scalar fields, and making some progress with analogous questions on nonminimally coupled fields.
First, we want to find a rotating generalization of the Wyman~II solutions, which would be, to our knowledge, the first example of a stationary --- but not static --- spacetime with a symmetry noninheriting, minimally coupled real scalar field.
Also, to complete this family of spacetimes, we shall generalize the static Wyman~II solutions to higher-dimensional spacetimes and investigate some three-dimensional Wyman-like solutions.
Finally, looking for the simplest candidate of symmetry noninheriting real scalar hair, we shall focus on stealth configurations, and prove an associated no-hair theorem for the Schwarzschild black hole. 

The paper is organized as follows.
In \cref{s:sym_inher} we briefly review the most important results on symmetry inheritance for minimally coupled scalar fields.
As a warm-up, in \cref{s:wymorig} we review Wyman's classification and derive again the Wyman~II solutions, introducing as well a new numerical global solution.
In \cref{s:rotwym} we spin up the Wyman~IIa solution using Hartle's perturbative formalism.
In \cref{s:nonmin} we prove the absence of nonminimally coupled stealth time-dependent real scalar fields on top of the Schwarzschild spacetime.  
Finally, we discuss our results and remaining open questions in \cref{s:discussion}.
We relegate the results on three and higher-dimensional spacetimes to the Appendices.
In \cref{a:3d} we derive both the static and stationary three-dimensional Wyman-like solutions, while in \cref{a:WymanD} we derive the static higher-dimensional generalization of the Wyman~II solutions.
Unless stated otherwise, throughout the paper we use natural units with $c = G = 1$.

%%%%%%%%%%%%%%%%%%%%%%%%%%%%%%%%%%%%%%%%%%%%%%%%%%%%%%%%%%%%%%%%%%%%%%%%%%%%%%%%%%%%
%%%%%%%%%%%%%%%%%%%%%%%%%%%%%%%%%%%%%%%%%%%%%%%%%%%%%%%%%%%%%%%%%%%%%%%%%%%%%%%%%%%%
\section{Symmetry inheritance of scalar fields revisited\label{s:sym_inher}} %%%%%%%
%%%%%%%%%%%%%%%%%%%%%%%%%%%%%%%%%%%%%%%%%%%%%%%%%%%%%%%%%%%%%%%%%%%%%%%%%%%%%%%%%%%%
%%%%%%%%%%%%%%%%%%%%%%%%%%%%%%%%%%%%%%%%%%%%%%%%%%%%%%%%%%%%%%%%%%%%%%%%%%%%%%%%%%%%

Let $(\mathcal{M},g_{ab})$ be a $D$-dimensional spacetime, with $D \ge 2$, solution of some gravitational field equations $E_{ab} = 8\pi T_{ab}$, where $E_{ab}$ is a generic smooth tensor field constructed from the spacetime metric, the Riemann tensor, the Levi-Civita tensor and covariant derivatives.
The matter content of this spacetime is that of a self-interacting real scalar field $\phi$, minimally coupled to the gravitational field, with energy-momentum tensor
\be\label{Ttensor}
T_{ab} = \nab{a}\phi \nab{b}\phi + \left(X - V\right) g_{ab}\,,
\ee
where $X \defeq -\frac{1}{2}\,\nab{c}\phi \nabla^c\phi$ is the kinetic term and $V=V(\phi)$ the scalar potential.
Being $T \defeq g^{ab} T_{ab}$ the trace of the energy-momentum tensor, it follows from \cref{Ttensor} that
\be
T &= (D-2)X - DV\,,\label{eq:T}\\
T_{ab} T^{ab} &= DX^2 + 2(2-D)XV + DV^2\,.\label{eq:TT}
\ee
The potential $V$ can be expressed in terms of these quantities as
\be\label{V}
V = -\frac{T}{D} \pm \frac{D-2}{2D} \, \sqrt{\frac{D T_{ab} T^{ab} - T^2}{D-1}}\,.
\ee
If we further assume that this spacetime admits a sufficiently smooth Killing vector field $K^a$, then $\Lie_K E_{ab} = 0$ and accordingly $\Lie_K T_{ab} = 0$. 
The Lie derivative of \cref{V} allows one to deduce~\cite{ISm15} the general properties of the Lie derivative $\Lie_K \phi$, provided a careful analysis of the points in which $V'(\phi) = 0$, e.g.\ for a massless scalar field with $V(\phi) = 0$.
The analysis in Ref.~\cite{ISm15}, however, contains a couple of minor technical gaps, i.e.\ the possible divergent points in the derivative of the square root in \cref{V}, and points with vanishing derivative $V'(\phi)$ in the $D=2$ case which require separate treatment.
Let us now show how to simplify and complete the analysis of the symmetry inheritance of scalar fields.

For $D \ge 3$, \cref{eq:T} can be solved for the kinetic term $X$,
\be\label{eq:X}
X = \frac{T+DV}{D-2}\,,
\ee
which inserted into \cref{eq:TT}, yields
\be\label{eq:newTT}
4D(D-1)V^2 + 8(D-1)TV + DT^2 - (D-2)^2 T_{ab} T^{ab} = 0\,,
\ee
and its Lie derivative with respect to $K^a$ reads
\be
(DV + T) \Lie_K V = 0\,.
\ee
On the open set where $DV + T \ne 0$, as well on the interior of the closed set where $DV + T = 0$ we immediately have $\Lie_K V = 0$ and by continuity this conclusion holds everywhere.
This latter equation means
\be\label{eq:LieV}
0 = \Lie_K V(\phi) = V'(\phi) \Lie_K \phi\,.
\ee
So, at any point where $V'(\phi) \ne 0$ symmetry inheritance follows immediately, i.e.\ $\Lie_K \phi = 0$.
Conversely, at any point where $V'(\phi) = 0$, the Lie derivative of \cref{eq:X} implies $\Lie_K X = 0$, which replaced in the Lie derivative of $T_{ab} K^b$ results
\be\label{eq:LieTK}
0 = \Lie_K (T_{ab} K^b) = (\Lie_K \phi) \, \nab{a} (\Lie_K \phi) = \frac{1}{2} \, \nab{a} (\Lie_K \phi)^2\,.
\ee
\Cref{eq:LieTK} states that within any open set invariant under the action of the Killing vector field $K^a$, in which $V'(\phi) = 0$, the Lie derivative $\Lie_K \phi$ is constant~\cite{ISm15}.
If the orbits of $K^a$ are compact then this is impossible unless $\Lie_K \phi = 0$. If the orbits are noncompact then the ``worst case scenario'' is that the scalar field grows linearly with the Killing parameter along the orbit --- as in the Wyman~II solutions.

For $D = 2$, \cref{eq:T} reduces to $V = -T/2$, which immediately leads to \cref{eq:LieV}. Thus, just as above, at any point where $V'(\phi) \ne 0$ the scalar field inherits the symmetries, $\Lie_K \phi = 0$.
Let $Z \subseteq \mathcal{M}$ be the set of points where $V'(\phi) = 0$. In any point of the interior $Z^\circ$, the Lie derivative of \cref{eq:TT} implies $\Lie_K X^2 = 0$.
Therefore, both in the open subset where $X \ne 0$, and in the interior of the subset where $X = 0$ we have $\Lie_K X = 0$. This result, by continuity, extends to $Z^\circ$ and to the whole set $Z$, which again brings us back to \cref{eq:LieTK}, with the same conclusions as above.

For the rest of the section we focus on general relativity, i.e.\ $E_{ab} = G_{ab}$, where $G_{ab} \defeq R_{ab}-\frac{1}{2}Rg_{ab}$ is the Einstein tensor.
For $D \ge 3$, the gravitational field equations may be put in the form
\be\label{EinsteinEqsRicci}
R_{ab} = 8\pi\left[\nab{a} \phi \nab{b} \phi + \frac{2}{D-2} \, V(\phi) g_{ab} \right].
\ee
The scalar field obeys the Klein--Gordon equation
\be
\Box\phi = V'(\phi)\,.
\ee
Note that a cosmological constant $\Lambda$ may be formally included as a constant term in the scalar potential $V$.

Symmetry noninheritance is ``easily destroyed'' by symmetric boundary conditions.
Consider, for example, a spacetime with a Killing horizon, generated by a Killing vector field $K^a$.
As $R_{ab} K^a K^b = 0$ on the horizon, it follows from the gravitational field equations that $\Lie_K \phi$ vanishes as well on the horizon and, since $\Lie_K \phi$ is constant under the assumptions stated above, the scalar field is symmetry inheriting.
This argument can be applied to black-hole horizons, as well as to cosmological horizons, and generalized even to a larger class of gravitational field equations~\cite{ISm17}.

Let us now investigate what the above general results imply on real scalar fields minimally coupled with general relativity, according to the number of spacetime dimensions.

The simplest, two-dimensional case is nearly trivial, as $G_{ab} = 0$, and the gravitational field equations become
\be\label{eom2d}
\nab{a}\phi \nab{b}\phi = \left( V - X \right) g_{ab}\,.
\ee
The trace of \cref{eom2d} gives $V = 0$, reducing the field equations to $\nab{a}\phi \nab{b}\phi = -X g_{ab}$, whose contraction with $\nabla^a\phi \nabla^b\phi$ further implies $X = 0$, so that $\nab{a}\phi \nab{b}\phi = 0$ and the scalar field is necessarily a constant. 

The three-dimensional case admits some nontrivial examples. For concreteness, assume that spacetime metric is stationary and axially symmetric, with corresponding commuting stationary Killing vector $k^a$ and axial Killing vector $m^a$.
In three spacetime dimensions, $k_a$ and $m_a$ automatically satisfy the Frobenius' condition,
\be
k_{[a} m_b \nab{c} k_{d]} = k_{[a} m_b \nab{c} m_{d]} = 0\,,
\ee 
and as a consequence the spacetime is Ricci-circular~\cite{Heusler}, i.e.\ $k_{[a} m_{b} R_{c]d} k^d = k_{[a} m_{b} R_{c]d} m^d = 0$.
Using a Ba\~nados--Teitelboim--Zanelli-like form of the metric with $(t,r,\varphi)$ coordinates, this means that $R_{tr} = R_{r\varphi} = 0$, so that the field equations, respectively, imply $\phi_t \phi_r = 0$ and $\phi_r \phi_\varphi = 0$.
The scalar field could either be symmetry inheriting, i.e.\ $\phi=\phi(r)$ or symmetry noninheriting, i.e.\ $\phi=\phi(t,\varphi)$.
In \cref{a:3d} we provide both the static and stationary solutions.

%%% memo: 0 = \Lie_m \omega_k = i_m \df\omega_k = -i_m {*(k \w R(k))} = {*(k \w m \w R(k))}

A similar argument applies for the four-dimensional case.
For definiteness, assume that the spacetime metric is stationary and axially symmetric, with  corresponding commuting stationary Killing vector $k^a$ and axial Killing vector $m^a$, which satisfy the Frobenius' condition.\footnote{Unlike in lower-dimensional cases, the Frobenius' condition is nontrivial in four spacetime dimensions.}
The spacetime is then Ricci-circular, which in Boyer--Lindquist-like coordinates $(t,r,\theta,\varphi)$ translates to $R_{tr} = R_{t\theta} = 0$.
Therefore, the most general form of the symmetry noninheriting real scalar field is $\phi = \gamma t + \psi(r,\theta)$, with some real constant $\gamma$ and some real function $\psi$, but Ricci-circular conditions and the field equations imply that either $\gamma = 0$ --- the symmetry inheriting case --- or $\psi$ is a constant, which may be removed by a redefinition of the coordinate $t$.
Furthermore, if $V(\phi) = 0$ and the spacetime metric is asymptotically flat, in the sense that $g^{tt} \sim -1 + O(r^{-\alpha})$ and $R \sim O(r^{-\beta})$ for some real positive constants $\alpha$ and $\beta$, then the trace of the Einstein field equations,
$R = 8\pi g^{tt} \gamma^2$, implies symmetry inheritance, i.e.\ $\gamma = 0$.
In conclusion, the form of the symmetry noninheriting scalar field is constrained to the simple expression $\phi = \gamma t$ and our quest will be first focused on horizonless spacetimes which are not asymptotically flat.

These results are easily generalized to higher-dimensional static spherically symmetric spacetimes as the angular part is ``integrated out'', i.e.\ we assume symmetry inheritance with respect to compact Killing orbits, and the analysis is reduced to the $t$-$r$ submanifold.
The generalization of the Wyman~II solutions to arbitrary spacetime dimensions are presented in \cref{a:WymanD}.

%%%%%%%%%%%%%%%%%%%%%%%%%%%%%%%%%%%%%%%%%%%%%%%%%%%%%%%%%%%%%
%%%%%%%%%%%%%%%%%%%%%%%%%%%%%%%%%%%%%%%%%%%%%%%%%%%%%%%%%%%%%
\section{Wyman's original solutions\label{s:wymorig}} %%%%%%%
%%%%%%%%%%%%%%%%%%%%%%%%%%%%%%%%%%%%%%%%%%%%%%%%%%%%%%%%%%%%%
%%%%%%%%%%%%%%%%%%%%%%%%%%%%%%%%%%%%%%%%%%%%%%%%%%%%%%%%%%%%%

In this section we review the Wyman classification of the static spherically symmetric solutions to the Einstein--Klein--Gordon equations for a real scalar field minimally coupled to Einstein gravity with no self-interacting potential.

Symmetry inheritance theorems state that $\phi = \gamma t + \psi(r)$, with some function~$\psi$ and constant $\gamma$.
The $tr$ component of the gravitational field equations implies $\phi_t\phi_r = \gamma \psi'(r) = 0$.
There exist two possible solutions of the latter equation,
\begin{description}
\item[$\gamma=0$] --- The scalar field inherits the spacetime symmetries, $\phi=\phi(r)$,
\item[$\psi=\const$] --- The scalar field grows linearly in time, $\phi = \gamma t$.
\end{description}

The first class of solutions, written in modern form~\cite{Virbhadra97}, reads
\be
\df s^2 = -f(r) \, \df t^2 + \frac{\df r^2}{f(r)} + \frac{r^2 - br}{f(r)} \, \df\Omega^2\,, \quad
\phi(r) = \frac{q}{b\sqrt{4\pi}} \, \ln f(r)\,,
\ee
where
\be
f(r) = \left( 1 - \frac{b}{r} \right)^{\!2m/b}\,, \quad
b = 2\sqrt{m^2 + q^2}\,,
\ee
and $m$ and $q$ are constants.

For the second class, starting with a general static, spherically symmetric background metric in Schwarzschild-like coordinates,
\be\label{ssmetric}
\df s^2 = -\e^{\nu(r)} \, \df t^2 + \e^{\lambda(r)}\,\df r^2 + r^2 \left(\df\theta^2 + \sin^2\theta\,\df\varphi^2\right),
\ee
the relevant field equations reduce to
\be\label{Wymansystem}
\lambda' - \nu' = \frac{2\left(1-\e^{\lambda}\right)}{r}\,,\quad
\lambda' + \nu' = 8\pi\gamma^2 r \e^{\lambda - \nu}\,,
\ee
where a prime denotes a derivative with respect to $r$ and from which one can obtain a second-order differential equation for $\lambda$
\be\label{eqlambda}
\lambda'' + \frac{3\left(\e^{\lambda}-1\right)}{r}\,\lambda' + \frac{2\left(\e^{\lambda}-2\right) \left(\e^{\lambda}-1\right)}{r^2} = 0\,.
\ee
Once the solution for $\lambda$ is known, $\nu$ can be obtained from the system~\eqref{Wymansystem}.

For this class of solutions, the nonvanishing components of the energy-momentum satisfy $-T^t_t = T^r_r = T^\theta_\theta = T^\varphi_\varphi = \gamma^2\e^{-\nu}/2$, hence one could interpret it as a specific perfect fluid for which the energy density and the pressure are identical --- see, however, Ref.~\cite{Madsen:1985}.

\Cref{eqlambda} admits a trivial global solution, $\lambda=\ln2$, which we refer to as Wyman~IIa,
\be\label{WymanIIa}
\lambda = \ln2\,,\quad \nu = \ln\left(8\pi\gamma^2 r^2\right).
\ee
In this coordinate system, the metric functions are always regular and have no zeros.
This solution represents a naked singularity, as can be seen from the diverging scalar invariants as $r\to0$, $R = -1/r^2$, $R_{ab} R^{ab} = 1/r^4$ and $R_{abcd} R^{abcd} = 3/r^4$.

Notice that \cref{eqlambda} also admits the solution $\e^\lambda = \gamma r/(\gamma r + \lambda_0)$ with some constant $\lambda_0$, which includes the trivial $\lambda=0$ solution. However, this solution is spurious and must be discarded, as the system~\eqref{Wymansystem} is no longer satisfied unless $\gamma=0$.

Around $r=0$, asking for regularity at the origin, \cref{eqlambda} can be solved as an approximate power-series solution, and we refer to as Wyman~IIb,
\begin{subequations}\label{WymanIIb}
\be
\lambda &= \frac{4\pi\gamma^2 r^2}{3} - \frac{56\pi^2\gamma^4 r^4}{45} + \frac{64\pi^3\gamma^6 r^6}{81} + O(r^8)\,,\\
\nu &= \frac{8\pi\gamma^2 r^2}{3} - \frac{64\pi^2\gamma^4 r^4}{45} + \frac{256\pi^3\gamma^6 r^6}{405} + O(r^8)\,.
\ee
\end{subequations}
For this approximate solution the curvature invariants are regular at $r=0$ and their values are $R = -8\pi \gamma^2$, $R_{ab} R^{ab} = 64\pi^2\gamma^4$ and $R_{abcd} R^{abcd} = 320\pi^2\gamma^4/3$.

Asymptotically, assuming that the solution of the system~\eqref{Wymansystem} can be expanded around the leading terms in \cref{WymanIIa}, the resulting subdominant terms introduce an oscillatory damped behaviour~\cite{Pi05,Jackiw06},
\begin{subequations}\label{WymanIIlarger}
\be
\lambda &\sim \ln2 + \frac{\alpha_\infty \sin \left(\sqrt{3} \ln \gamma r\right) + \beta_\infty \cos \left(\sqrt{3} \ln \gamma r\right)}{r}\,,\label{lambdalarger}\\
\nu &\sim \ln\left(8\pi\gamma^2 r^2\right) - \sqrt{3}\,\frac{\alpha_\infty \cos \left(\sqrt{3} \ln \gamma r\right) - \beta_\infty \sin \left(\sqrt{3} \ln \gamma r\right)}{r}\,,
\ee
\end{subequations}
where $\alpha_\infty$ and $\beta_\infty$ are integration constants.

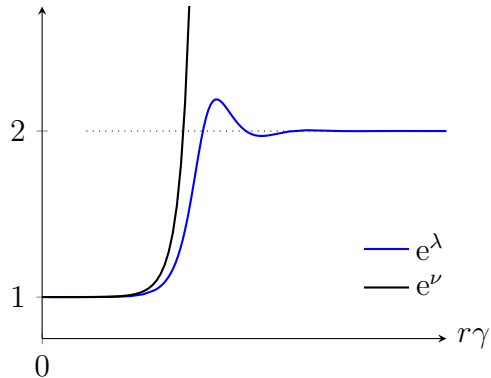
\begin{figure}[htb]
\centering
\begin{tikzpicture}
\begin{axis} [
  ymin=0.75,ymax=2.75,
  height=6cm,
  axis lines=left,
  xtick={-6.74},
  ytick={1,2},
  xticklabels={0},
  xlabel=$r\gamma$,
  every axis x label/.style={at={(ticklabel* cs:1.0)}, anchor=west},
  every axis y label/.style={at={(ticklabel* cs:1.0)}, anchor=south},
  legend style={anchor=north, at={(0.9,0.35)}, draw=none},
  ]

% Exp(lambda)
\addplot[thick, color=blue]
  coordinates {(-6.44724, 1.00001) (-5.98672, 1.00003) (-5.5262, 
    1.00007) (-5.06569, 1.00017) (-4.60517, 1.00042) (-4.14465, 
    1.00105) (-3.68414, 1.00264) (-3.22362, 1.00663) (-2.7631, 
    1.01662) (-2.30259, 1.04152) (-2.21048, 1.04983) (-2.11838, 
    1.05977) (-2.02627, 1.07166) (-1.93417, 1.08586) (-1.84207, 
    1.10279) (-1.74996, 1.12294) (-1.65786, 1.14685) (-1.56576, 
    1.17512) (-1.47365, 1.20843) (-1.38155, 1.24743) (-1.28945, 
    1.29278) (-1.19734, 1.34504) (-1.10524, 1.40457) (-1.01314, 
    1.4714) (-0.921034, 1.54508) (-0.828931, 1.62452) (-0.736827, 
    1.7079) (-0.644724, 1.79266) (-0.55262, 1.87565) (-0.460517, 
    1.95348) (-0.368414, 2.02295) (-0.27631, 2.08148) (-0.184207, 
    2.12747) (-0.0921034, 2.16044) (0., 2.1809) (0.0921034, 
    2.19018) (0.184207, 2.19006) (0.27631, 2.18253) (0.368414, 
    2.16952) (0.460517, 2.1528) (0.55262, 2.13386) (0.644724, 
    2.11392) (0.736827, 2.09393) (0.828931, 2.07462) (0.921034, 
    2.05649) (1.01314, 2.03989) (1.10524, 2.02505) (1.19734, 
    2.01207) (1.28945, 2.001) (1.38155, 1.99181) (1.47365, 
    1.98443) (1.56576, 1.97875) (1.65786, 1.97466) (1.74996, 
    1.97198) (1.84207, 1.97058) (1.93417, 1.97028) (2.02627, 
    1.9709) (2.11838, 1.97229) (2.21048, 1.97429) (2.30259, 
    1.97672) (2.99336, 1.99715) (3.68414, 2.00486) (4.37491, 
    2.00248) (5.06569, 1.99969) (5.75646, 1.99926) (6.44724, 
    1.99981) (7.13801, 2.00011) (7.82879, 2.00009) (8.51956, 
    2.) (9.21034, 1.99998)};

% Exp(nu)
\addplot[thick,color=black]
  coordinates {(-6.74657, 1.00001) (-6.58539, 1.00002) (-6.42421, 
    1.00002) (-6.26303, 1.00003) (-6.10185, 1.00004) (-5.94067, 
    1.00006) (-5.77949, 1.00008) (-5.61831, 1.00011) (-5.45713, 
    1.00015) (-5.29595, 1.00021) (-5.13476, 1.00029) (-4.97358, 
    1.0004) (-4.8124, 1.00055) (-4.65122, 1.00076) (-4.49004, 
    1.00105) (-4.32886, 1.00146) (-4.16768, 1.00201) (-4.0065, 
    1.00278) (-3.84532, 1.00383) (-3.68414, 1.00529) (-3.52296, 
    1.00731) (-3.36177, 1.0101) (-3.20059, 1.01396) (-3.03941, 
    1.0193) (-2.87823, 1.0267) (-2.71705, 1.03697) (-2.55587, 
    1.05124) (-2.39469, 1.07114) (-2.23351, 1.09896) (-2.07233, 
    1.13806) (-1.91115, 1.19335) (-1.74996, 1.27217) (-1.58878, 
    1.38571) (-1.4276, 1.55139) (-1.26642, 1.79686) (-1.10524, 
    2.16676) (-0.94406, 2.73362) (-0.782879, 3.61468) (-0.621698, 
    4.99524) (-0.460517, 7.15652) (-0.299336, 10.5032) (-0.138155, 
    15.5859) (0.0230259, 23.1205) (0.184207, 34.0162) (0.345388, 
    49.4263) (0.506569, 70.8363) (0.66775, 100.201)};

\addplot [dotted] {2};

\legend{$\e^\lambda$,$\e^\nu$}
\end{axis}
\end{tikzpicture}
\caption{Numerical solution of the system~\eqref{Wymansystem}. This global nontrivial solution interpolates between the Wyman~IIa and IIb solutions.}\label{fig:numsol}
\end{figure}

In general, \cref{eqlambda} can be integrated numerically. Using as boundary conditions \cref{WymanIIb}, the numerical solution for $\lambda$, together with the corresponding solution for $\nu$, are shown in \cref{fig:numsol}.
We notice that this solution ``interpolates'' between the Wyman~IIb (as $r\to0$) and Wyman~IIa (as $r\to\infty$) solutions.
The metric function $\e^\lambda$ manifests a global maximum for $r\gamma\approx1.14564$ and then oscillates damped to reach asymptotically the value~$2$ in agreement with \cref{lambdalarger}.
The metric function $\e^\nu$ asymptotes rapidly to $8\pi\gamma^2 r^2$.
The metric functions and the curvature invariants are regular everywhere, but the solution cannot be interpreted as a soliton as it does not have finite energy.

%%%%%%%%%%%%%%%%%%%%%%%%%%%%%%%%%%%%%%%%%%%%%%%%%%%%%%%%%%%%%%%%%%%%
%%%%%%%%%%%%%%%%%%%%%%%%%%%%%%%%%%%%%%%%%%%%%%%%%%%%%%%%%%%%%%%%%%%%
\section{Spinning up the Wyman~IIa solution\label{s:rotwym}} %%%%%%%
%%%%%%%%%%%%%%%%%%%%%%%%%%%%%%%%%%%%%%%%%%%%%%%%%%%%%%%%%%%%%%%%%%%%
%%%%%%%%%%%%%%%%%%%%%%%%%%%%%%%%%%%%%%%%%%%%%%%%%%%%%%%%%%%%%%%%%%%%

The generalization from static to stationary solutions is rarely an easy task.
A popular technique to generate rotating solutions starting from static metrics is the Newman--Janis algorithm~\cite{Newman:1965tw}, which has been successfully applied to a number of cases, although its feasibility for the symmetry inheriting Fisher solution~\cite{Krori:1982,Agnese:1985xj} turned out not correct~\cite{Bogush:2020lkp}.
Since we are interested in symmetry noninheriting solutions, the form of the scalar field is the same even for rotating spacetimes, i.e.\ $\phi = \gamma t$.
This fact yields that the metric obtained with the Newman--Janis algorithm does not satisfy the equations of motion --- not even at linear order in the rotation parameter. 
Moreover, as discussed in \cref{s:sym_inher}, if a rotating symmetry noninheriting solution exists, this must be horizonless and nonasymptotically flat. These properties limit the physical interest for an exact solution and thus we decided to investigate the existence of a rotating extension of the Wyman~II solutions within a perturbative framework.

Following Hartle~\cite{Hartle67}, our starting point is the slow-rotation expansion around the Wyman~IIa background solution, including up to second-order spin corrections
\be
\df s^2 = &- 8\pi\gamma^2 r^2 \left(1 + 2\epsilon^2\left[h_0(r) + h_2(r) P_2(\theta)\right] - \frac{\epsilon^2 \omega(r)^2 \sin^2\theta}{8\pi\gamma^2}\right) \df t^2\0\\
&- 2\epsilon r^2 \sin^2\theta \omega(r)\,\df t\df\varphi + 2\left(1 + \frac{4\epsilon^2}{r}\left[m_0(r) + m_2(r) P_2(\theta)\right]\right)\df r^2\0\\
&+ r^2 \left[1 + 2\epsilon^2 k_2(r) P_2(\theta)\right] \left(\df\theta^2 + \sin^2\theta\,\df\varphi^2\right)
+ O(\epsilon^3)\,,
\ee
where $\epsilon$ is a book-keeping parameter, $P_2(\theta) = (3\cos^2\theta - 1)/2$, and $h_0$, $h_2$, $\omega$, $m_0$, $m_2$ and $k_2$ are functions of the radial coordinate to be determined.
We now solve the field equations order by order in $\epsilon$ and we require that, for $r\to\infty$, the perturbative terms in the metric functions are at most constant.

The Einstein--Klein--Gordon equations at $O(\epsilon)$ require
\be
\omega'' + \frac{3 \omega'}{r} = 0\,,
\ee
whose general solution is $\omega = \beta + \alpha/r^2$. However, the $\beta$-term would modify the asymptotic behaviour of the $g_{t\varphi}$ component of the metric, i.e.\ $g_{t\varphi}\sim -\beta\epsilon r^2\sin^2\theta$, and to avoid this to happen we set $\beta=0$.

The Einstein--Klein--Gordon equations at $O(\epsilon^2)$ are more involved.
From the $r\theta$ component we get the differential relation
\be\label{m2diffrelation}
m_2 = \frac{r^2}{4}\left(h_2' + k_2'\right),
\ee
which substituted in the difference between the $\theta\theta$ and $\varphi\varphi$ components leads to
\be\label{k2primerelation}
k_2' = \frac{\alpha^2}{12\pi\gamma^2 r^5} - h_2' - \frac{2 h_2}{r}\,.
\ee
Combining \cref{m2diffrelation,k2primerelation} we get the algebraic relation
\be\label{m2relation}
m_2 = \frac{\alpha^2}{48 \pi \gamma^2 r^3}-\frac{r}{2} h_2\,.
\ee

Substituting in the remaining equations, the $rr$ component is satisfied for any value of $\theta$ for
\be
m_0 &= \frac{\alpha^2}{144\pi\gamma^2 r^3} + \frac{1}{6} r \left(r h_0' + h_0\right),\label{m0rule}\\
k_2 &= -\frac{1}{4} \left(r h_2' + 6 h_2\right).\label{k2rule}
\ee
The $\theta\theta$ component gives a decoupled equation for $h_0$ whose solution is
\be
h_0 &= \frac{\alpha^2}{48 \pi \gamma^2 r^4} + \frac{c_1 \sin \left(\sqrt{3} \ln \gamma r\right)+c_2 \cos \left(\sqrt{3} \ln \gamma r\right)}{r}\,,
\ee
which substituted into the $tt$ component gives an equation for $h_2$ whose solution is
\be\label{h2sol}
h_2 &= c_3 r^2 + \frac{\alpha^2 (1+ 6\ln \gamma r)}{108 \pi  \gamma^2 r^4} + \frac{c_4}{r^4}\,.
\ee
Notice that, when solving for $h_0$ and $h_2$, we have exercised the freedom of choice on the integration constants to have a dimensionless argument of the logarithm.
Finally, using \cref{m2relation,m0rule,k2primerelation} we get
\be
m_2 &= -\frac{1}{2} c_3 r^3 + \frac{\alpha^2 (7-12 \ln \gamma r)}{432\pi\gamma^2 r^3} - \frac{c_4}{2 r^3}\,,\\
m_0 &= -\frac{\alpha^2}{288\pi\gamma^2 r^3} + \frac{c_1 \cos \left(\sqrt{3} \ln \gamma r\right)-c_2 \sin \left(\sqrt{3} \ln \gamma r\right)}{2 \sqrt{3}}\,,\\
k_2 &= - 2 c_3 r^2 - \frac{\alpha^2 (2 + 3\ln \gamma r)}{108\pi\gamma^2 r^4} -\frac{c_4}{2 r^4}\,.
\ee
Many of the integration constants can be set to zero by the following arguments.
The term proportional to $r^2$ in \cref{h2sol} would change the asymptotic behaviour of the static background solution, i.e.\ $g_{tt}\sim c_3 r^4$, and hence we set $c_3=0$.
To avoid oscillatory behaviour at large $r$ for $m_0$ we need to set $c_1=c_2=0$ as well --- differently to what happens for $h_0$ and for the global Wyman~II solution, cfr.\ \cref{WymanIIlarger}, the oscillatory behaviour is not damped for $m_2$. It is interesting to notice that the peculiar functional form of the asymptotic oscillatory behaviour for the global Wyman~II solution is somehow ``unveiled'' by rotation.
In conclusion, the first-order correction introduces two integration constants $\alpha$ and $\beta$, while the second-order correction introduces four new integration constants $c_1$, $c_2$, $c_3$ and $c_4$. However, our assumptions on the perturbative metric functions further imply $\beta=c_1=c_2=c_3=0$ and we are left with two new integration constants.
This solution is still a (rotating) naked singularity and the curvature invariants get a $O(\epsilon^2)$ correction to the background values reported below \cref{WymanIIa}.

In order to add some physical interpretation to this solution, we may --- at least formally --- find associated charges.
In a spacetime with isometries, the most natural choice is to use the Komar formulae for the mass $M_{\mathcal{S}}$ and angular momentum $J_{\mathcal{S}}$, given in terms of integrals of the Hodge duals of the 2-forms $(\df k)_{ab} = 2\nab{[a}k_{b]}$ and $(\df m)_{ab} = 2\nab{[a}m_{b]}$, over a closed 2-surface $\mathcal{S}$~\cite{Heusler},
\be
M_{\mathcal{S}} = -\frac{1}{8\pi} \oint_{\mathcal{S}} {\Hodge\,\df k}\,, \quad J_{\mathcal{S}} = \frac{1}{16\pi} \oint_{\mathcal{S}} {\Hodge\,\df m}\,.
\ee
For the perturbative rotating Wyman~IIa solution, these two integrals evaluated on a 2-sphere $S^2$, defined by constant $(t,r)$, give
\be
M_{S^2} = 2\gamma\sqrt{\pi}\,r^2 + \frac{\alpha^2}{18\gamma\sqrt{\pi} r^2} \, \epsilon^2 + O(\epsilon^3)\,,
\ee
and
\be
J_{S^2} = \frac{\alpha}{12\gamma\sqrt{\pi}}\,\epsilon + O(\epsilon^3)\,.
\ee

On the one hand, the Komar mass diverges as $r \to \infty$, which is not surprising as the metric is not asymptotically flat --- note, however, that the $O(\epsilon^2)$ term vanishes as $r \to \infty$.
On the other hand, the Komar angular momentum is well-defined and may be used for a physical interpretation of the perturbation parameter $\epsilon$.
Unfortunately, it is not quite clear how to give any meaningful definition of scalar charge $Q_\phi$ that could be associated with this solution.

%%%%%%%%%%%%%%%%%%%%%%%%%%%%%%%%%%%%%%%%%%%%%%%%%%%%%%%%%%%%%%%%%%%%%%%%%%%%%%%%%%%%%%%%%%%%%%%%%%%%%%%%%
%%%%%%%%%%%%%%%%%%%%%%%%%%%%%%%%%%%%%%%%%%%%%%%%%%%%%%%%%%%%%%%%%%%%%%%%%%%%%%%%%%%%%%%%%%%%%%%%%%%%%%%%%
\section{Is there a symmetry noninheriting hair with nonminimal coupling?\label{s:nonmin}} %%%%%%%%%%%%%%
%%%%%%%%%%%%%%%%%%%%%%%%%%%%%%%%%%%%%%%%%%%%%%%%%%%%%%%%%%%%%%%%%%%%%%%%%%%%%%%%%%%%%%%%%%%%%%%%%%%%%%%%%
%%%%%%%%%%%%%%%%%%%%%%%%%%%%%%%%%%%%%%%%%%%%%%%%%%%%%%%%%%%%%%%%%%%%%%%%%%%%%%%%%%%%%%%%%%%%%%%%%%%%%%%%%

Wyman's original solutions and the rotating extension derived above serve as a proof-of-principle, but can hardly be seen as valuable physical examples since they exhibit some pathological features --- e.g.\ Wyman~IIa contains a naked singularity.

Minimally coupled real scalar fields cannot evade no-hair theorems via symmetry noninheritance, but the landscape of gravitational theories with scalar fields is vast, with a plethora of models and solutions~\cite{Kobayashi19,Faraoni:2021nhi}. Some of these theories admit static black-hole solutions with symmetry noninheriting time-dependent scalar fields, see e.g.\ Ref.~\cite{Babichev:2013cya}. Yet, for the simplest nonminimal coupling, defined below, the question about the existence of a symmetry noninheriting hair is still open.

In this section we consider a nonself-interacting scalar field nonminimally coupled to gravitation, described by the action
\be\label{nonmintheory}
S = \int \df^4 x \, \sqrt{-g} \, \left( \frac{1}{16\pi}\,R + X - f(\phi)R \right).
\ee
The function $f$ defines the nonminimal coupling and enters in the gravitational and generalized Klein--Gordon equations,
\be
\big[1 - 16\pi f(\phi)\big] G_{ab} & = 8\pi T_{ab}^{(\phi)}\,,\\
\Box\phi & = f'(\phi)R\,,
\ee
where the energy-momentum tensor is
\be
T_{ab}^{(\phi)} = \big[1 - 2f''(\phi)\big] \nab{a}\phi\nab{b}\phi - 2f'(\phi)\,\nab{a}\nab{b}\phi + \big[\left(1-4f''(\phi)\right)X + 2f'(\phi)\,\Box\phi \big] g_{ab}\,.
\ee

For the rest of the section we focus on the simplest choice for the function $f$, that gives the only dimension-four operator that accounts for nonminimal coupling \cite{Jordan:1959eg,Adler:1982ri},
\be
f(\phi) = \frac{1}{2}\,\xi\phi^2\,,
\ee
with some real parameter $\xi \ne 0$. For the particular value $\xi_c = 1/6$, the field equations are conformally invariant~\cite{Wald}.

The first attempt of a systematic treatment of symmetry inheritance for nonminimally coupled scalar fields~\cite{BS17} points to the so-called stealth configurations, i.e.\ those with vanishing energy-momentum tensor, as first candidates for symmetry noninheriting scalar black-hole hair.
This is motivated by the existence of a time-dependent stealth real scalar field on top of the Minkowski spacetime~\cite{ABMTZ05} and in what follows we investigate the extension of these configurations to spacetimes with $M>0$, i.e.\ on top of Schwarzschild black holes, described by the line element
\be
\df s^2 = -\left(1-\frac{2M}{r}\right)\df t^2 + \left(1-\frac{2M}{r}\right)^{-1}\df r^2 + r^2 \, \df\Omega^2\,.
\ee

The nonminimally coupled theory \eqref{nonmintheory} can be turned into a theory with a minimally coupled scalar field, using a $\phi$-dependent conformal transformation \cite{Saa96a}. However, there is no \emph{a priori} reason that the transformed scalar field $\tilde\phi$ will still be stealth --- which would immediately lead to the conclusion that $\tilde\phi$ is trivial~\cite{Sokolowski03,BS17} --- nor that the transformed spacetime metric inherits the isometries of the original metric --- as the scalar field is, by assumption, symmetry noninheriting. Hence, we have to look more carefully into the field equations.

Let us denote by $S \subseteq \mathcal{M}$ the closed set of points where $\phi_t = 0$ and assume that the domain of the problem is its complement, the open set $\mathcal{M} - S$. The field equations for stealth configurations are reduced to 
\be
0 & = \Box\phi\,,\\
0 & = \left(1-2\xi\right) \nab{a}\phi \nab{b}\phi - 2\xi\phi \nab{a}\nab{b}\phi + \left(1-4\xi\right) X g_{ab}\,.\label{eq:stealth}
\ee
The $t\theta$ and $t\varphi$ components of \cref{eq:stealth} read
\be
\left(1 - 2\xi\right) \phi_t \, \phi_i = 2\xi \phi \, \phi_{ti}\,,
\ee
for $i \in \{\theta,\varphi\}$, which can be written as
\be
\dd_i \left[ 2\xi \ln(\phi\,\phi_t) - \ln\phi \right] = 0\,,
\ee
so that $2\xi \ln(\phi\,\phi_t) - \ln\phi$ is some function which depends on $t$ and $r$ only. Then $\phi^{2\xi - 1} (\phi_t)^{2\xi}$ is a function of $t$ and $r$ and
\be\label{ansatzphi}
\phi(t,r,\theta,\varphi) =
\begin{cases}
\big[\alpha(t,r) + \beta(r,\theta,\varphi) \big]^{\frac{2\xi}{4\xi - 1}} & \text{ if } \xi \ne 1/4 \\
\alpha(t,r)\,\beta(r,\theta,\varphi) & \text{ if } \xi = 1/4 
\end{cases}\,,
\ee
for some functions $\alpha$ and $\beta$ to be determined.

Suppose that $\xi \ne 1/4$ and insert the Ansatz~\eqref{ansatzphi} back into \cref{eq:stealth}. The $tr$ equation can be solved for $\alpha$,
\be
\alpha(t,r) = u(t) \sqrt{1 - \frac{2M}{r}} + v(r)\,,
\ee
with some functions $u(t)$ and $v(r)$. The $r\theta$ and $r\varphi$ equations give $\dd_i (r \beta_r - \beta) = 0$ with $i \in \{\theta,\varphi\}$, so that $\beta = \beta(r)$. In other words,
\be\label{phisolxi}
\phi(t,r,\theta,\varphi) = \left( u(t) \sqrt{1 - \frac{2M}{r}} + v(r) + \beta(r) \right)^{\!\frac{2\xi}{4\xi - 1}}\,.
\ee
The second time derivative of the $\theta\theta$ component of \cref{eq:stealth} once we substitute the solution~\eqref{phisolxi}, taking into account that $\dot{u} \ne 0$ on the set $\mathcal{M}-S$, where a dot denotes a derivative with respect to $t$, reads
\be
r^4 \dddot{u} - M\left(5M-2r\right)\dot{u} = 0\,.
\ee
This is a contradiction unless $M = 0$, leading us back to the trivial Minkowski case.

The $\xi = 1/4$ case is analogous. The $tr$ equation can be solved again for $\alpha$,
\be
\alpha(t,r) = \exp\left( u(t) \sqrt{1 - \frac{2M}{r}} + v(r) \right).
\ee
with some functions $u(t)$ and $v(r)$. When this solution is substituted in the $tt$ component of \cref{eq:stealth}, it implies that
\be
M^2\beta u - r^4 \beta \ddot{u} + M\sqrt{r^3(r-2M)} \, \beta_r = 0\,,
\ee
whose time derivative leads to
\be
\left(r^4 \dddot{u} - M^2 \dot{u}\right)\beta = 0\,,
\ee
again, a contradiction unless $M = 0$.

In conclusion, we have eliminated the possibility of a time-dependent stealth scalar field on top of the Schwarzschild black hole.

%%%%%%%%%%%%%%%%%%%%%%%%%%%%%%%%%%%%%%%%%%%%%%%%%%%%%%%%%%%%
%%%%%%%%%%%%%%%%%%%%%%%%%%%%%%%%%%%%%%%%%%%%%%%%%%%%%%%%%%%%
\section{Discussion\label{s:discussion}} %%%%%%%%%%%%%%%%%%%
%%%%%%%%%%%%%%%%%%%%%%%%%%%%%%%%%%%%%%%%%%%%%%%%%%%%%%%%%%%%
%%%%%%%%%%%%%%%%%%%%%%%%%%%%%%%%%%%%%%%%%%%%%%%%%%%%%%%%%%%%

This paper filled some missing gaps in understanding the symmetry inheritance properties of real scalar fields.

After an exhaustive overview of the original Wyman~II solutions, that we have also extended to three and higher dimensions, one of the novel results is the slowly-rotating version of the Wyman~IIa solution which is, to our knowledge, the first example of an exact, nonstatic solution --- albeit perturbative --- with symmetry noninheriting real scalar field.
Even if this solution describes a rotating naked singularity, and therefore it is of little interest for astrophysics, it represents an important proof-of-principle about the existence of rotating configurations sourced by symmetry noninheriting real scalar fields.
Moreover, it completes the set of possibilities within the family of minimally coupled real scalar fields --- with the strong-field rotating solution being the (possible) only missing piece. The existence of such an exact rotating Wyman II solution together with the boundary conditions or constraints on global charges for its uniqueness are still open questions. Incidentally, we notice that this possible exact solution might not admit our perturbative solution as a limit: in our derivation we have assumed the perturbations to be subdominant with respect to the background Wyman IIa solution, but in principle this is not guaranteed in the strong-field regime, because of the absence of asymptotic boundary conditions.

Having systematically surveyed minimally coupled real scalar fields and shown that they cannot evade no-hair theorems via symmetry noninheritance, we have made one important step ahead on the front of nonminimally coupled real scalar fields. We have explicitly proven that there cannot exist any time-dependent stealth real scalar field on top of the Schwarzschild black hole.
We expect this no-hair result to hold also for Kerr black holes, but we leave this question for future work. From a broader point of view, it is not yet quite clear whether different forms of the nonminimal coupling function $f(\phi)$ might admit evasion of no-hair theorems via symmetry noninheritance.

%%%%%%%%%%%%%%%%%%%%%%%%%%%%%%%%%%%%%%%%%%%%%%%%%%%%%%%%%%
%%%%%%%%%%%%%%%%%%%%%%%%%%%%%%%%%%%%%%%%%%%%%%%%%%%%%%%%%%
\ack
EF acknowledges partial financial support by CNPq Brazil, process nos.\ 302397/2019-1, 300328/2020-6, 301088/2020-9 and by the Italian Ministry of Education and Scientific Research (MIUR) under the grant PRIN MIUR 2017-MB8AEZ\@.

%%%%%%%%%%%%%%%%%%%%%%%%%%%%%%%%%%%%%%%%%%%%%%%%%%%%%%%%%%
%%%%%%%%%%%%%%%%%%%%%%%%%%%%%%%%%%%%%%%%%%%%%%%%%%%%%%%%%%

%%%%%%%%%%%%%%%%%%%%%%%%%%%%%%%%%%%%%%%%%%%%%%%%%%%%%%%%%%
%%%%%%%%%%%%%%%%%%%%%%%%%%%%%%%%%%%%%%%%%%%%%%%%%%%%%%%%%%
\appendix
%%%%%%%%%%%%%%%%%%%%%%%%%%%%%%%%%%%%%%%%%%%%%%%%%%%%%%%%%%
%%%%%%%%%%%%%%%%%%%%%%%%%%%%%%%%%%%%%%%%%%%%%%%%%%%%%%%%%%

%%%%%%%%%%%%%%%%%%%%%%%%%%%%%%%%%%%%%%%%%%%%%%%%%%%%%%%%%%%%%%%%%%%%%%%
%%%%%%%%%%%%%%%%%%%%%%%%%%%%%%%%%%%%%%%%%%%%%%%%%%%%%%%%%%%%%%%%%%%%%%%
\section{Static and rotating solutions in three dimensions\label{a:3d}}
%%%%%%%%%%%%%%%%%%%%%%%%%%%%%%%%%%%%%%%%%%%%%%%%%%%%%%%%%%%%%%%%%%%%%%%
%%%%%%%%%%%%%%%%%%%%%%%%%%%%%%%%%%%%%%%%%%%%%%%%%%%%%%%%%%%%%%%%%%%%%%%

In three spacetime dimensions the Einstein--Klein--Gordon equations can be solved analytically both for the static and for the axisymmetric case.
Let us begin with a Ba\~nados--Teitelboim--Zanelli-like ansatz for the metric and a general real scalar field
\be%
\df s^2 = - \e^{\nu(r)}\,\df t^2 + \e^{\lambda(r)}\,\df r^2 + r^2 \left(\df\varphi + \e^{\Gamma(r)}\,\df t\right)^2,\quad
\phi=\phi(t,r,\varphi)\,.
\ee%
We know that symmetry noninheritance further constraints the scalar field not to be a function of $r$, and in this case, the Klein--Gordon equation reads
\be
\left(\frac{1}{r^2}-\e^{2\Gamma-\nu}\right) \phi_{\varphi\varphi} + \e^{-\nu} \left(2 \e^\Gamma \phi_{t\varphi} - \phi_{tt} \right) = 0\,,
\ee
which admits a simple solution $\phi = \gamma t + \alpha \varphi$.
The requirement $\phi|_{\varphi=0} = \phi|_{\varphi=2\pi}$ implies $\alpha=0$.
The $tr$ component of the Einstein field equations is
\be\label{eq13}
\Gamma'' + \Gamma' \left(\frac{3}{r} - \frac{\lambda'+\nu'}{2}\right) + \Gamma'^2 = 0\,,
\ee
which admits a constant solution, i.e.\ $\Gamma=\ln \chi$.
The remaining field equations reduce to:
\be
\nu' = \lambda'\,,\quad
16 \pi \gamma^2 r \e^{-\nu} = \e^{-\lambda}\left( \lambda' + \nu'\right)\,,\quad
16 \pi \gamma^2 \e^{\lambda - \nu} = 2\nu'' + \nu'^2 - \lambda' \nu'\,.
\ee

Combining the first two equations we get $\lambda = \nu+c_1$ and $\nu = 4 \pi \gamma^2 c_1 r^2 + c_2$, and the third equation is identically satisfied.

Finally, the solution is
\be\label{wyman3d}
\df s^2 = - e^{4 \pi  c_1 \gamma^2 r^2+c_2}\,\df t^2 + c_1 e^{4 \pi  c_1 \gamma^2 r^2+c_2}\,\df r^2 + r^2\left(\chi\,\df t + \df\varphi\right)^2\,,\quad
\phi = \gamma t\,.
\ee
A convenient gauge choice would be $c_1=1$ and $c_2=0$.

Both the metric components and the curvature invariants are regular at $r=0$, i.e., $R = -8 \pi \e^{-c_2} \gamma^2, R_{ab}R^{ab} = 64 \pi^2 \e^{-2c_2} \gamma^4, R_{abcd}R^{abcd} = 192 \pi^2 \e^{-2c_2} \gamma^4$.
The nonrotating solution (i.e.\ $\e^{\Gamma(r)}=0$) is easily obtained as the $\chi\to0$ limit of \cref{wyman3d}.

%%%%%%%%%%%%%%%%%%%%%%%%%%%%%%%%%%%%%%%%%%%%%%%%%%%%%%%%%%%%%%%%%%%%%%%%%%%%%%%%%%%%%%%%%
%%%%%%%%%%%%%%%%%%%%%%%%%%%%%%%%%%%%%%%%%%%%%%%%%%%%%%%%%%%%%%%%%%%%%%%%%%%%%%%%%%%%%%%%%
\section{Higher-dimensional generalization of the Wyman~II solutions\label{a:WymanD}} %%%%%%
%%%%%%%%%%%%%%%%%%%%%%%%%%%%%%%%%%%%%%%%%%%%%%%%%%%%%%%%%%%%%%%%%%%%%%%%%%%%%%%%%%%%%%%%%
%%%%%%%%%%%%%%%%%%%%%%%%%%%%%%%%%%%%%%%%%%%%%%%%%%%%%%%%%%%%%%%%%%%%%%%%%%%%%%%%%%%%%%%%%

The Wyman~IIa and IIb solutions presented in \cref{s:wymorig} can be easily generalized to arbitrary $D>3$ spacetime dimensions.

Consider a static spherically symmetric $D$-dimensional spacetime
\be
\df s^2 = -\e^{\nu(r)} \df t^2 + \e^{\lambda(r)} \df r^2 + r^2 \df\Omega_{D-2}^2\,,
\ee
in coordinates $x^0 = t$, $x^1 = r$, $x^i = \theta^i$ for $i \in \{2,\dots,D-1\}$;
the angular coordinates are defined such that $0 \le \theta^i \le \pi$ for $i \in \{2,\dots,D-2\}$ and $0 \le \theta^{D-1} \le 2\pi$.
Using the Appendix of Ref.~\cite{Bonora:2011mf}, the angular components of the metric can be written as $g_{ii} = r^2 \Pi(i)$ where
\be
\Pi(i) =
\begin{cases}
1\,, & i = 2\\
\prod_{m=2}^{i-1} \sin^2\theta^m\,, & i \ge 3
\end{cases}\,.
\ee

The Klein--Gordon equation is identically satisfied for $\phi=\gamma t$ while the nonzero components of the gravitational field equations written as in \cref{EinsteinEqsRicci} read
\begin{subequations}\be
R_{tt} &= \frac{\e^{\nu - \lambda}}{4} \left( 2\nu'' + \nu'^2 - \nu'\lambda' + 2\,\frac{D-2}{r}\,\nu' \right) = 8\pi\gamma^2\,,\\
R_{rr} &= -\frac{1}{4} \left( 2\nu'' + \nu'^2 - \nu'\lambda' - 2\,\frac{D-2}{r}\,\lambda' \right) = 0\,,\\
R_{ii} &= \frac{\Pi(i)}{2} \left[ r \e^{-\lambda} \left(\lambda' - \nu'\right) + 2(D-3)\left(1 - \e^{-\lambda}\right) \right] = 0\,.
\ee\end{subequations}

The difference of the first two equations together with the third equation are the $D$-dimensional generalization of \cref{Wymansystem}
\be\label{WymansystemD}
\lambda' + \nu' = 8\pi\gamma^2 \, \frac{2}{D-2} \, r \e^{\lambda - \nu}\,,\quad
\lambda' - \nu' = 2\,\frac{D-3}{r} \left(1 - \e^\lambda\right),
\ee
that can be combined to get the $D$-dimensional generalization of \cref{eqlambda}
\be
\lambda'' + \frac{3(D-3)\e^\lambda - (2D-5)}{r}\,\lambda' + 2(D-3)\,\frac{(\e^\lambda - 1)\left[(D-3)\e^\lambda - (D-2)\right]}{r^2} = 0\,.
\ee

Finally, the $D$-dimensional Wyman~IIa solution is
\be
\nu = \ln\left(\frac{8\pi\gamma^2 r^2}{D-3} \right), \quad \lambda =  \ln\left(\frac{D-2}{D-3} \right).
\ee

A direct computation reveals that the Wyman~IIb solution can be generalized as well. Around $r=0$, asking for regularity at the origin, \cref{WymansystemD} can be solved as an approximate power-series solution
\begin{subequations}\be
\lambda &= \frac{8 \pi  \gamma^2 r^2}{(D-2) (D-1)} - \frac{32 \pi^2 (D-3) (2D-1) \gamma^4 r^4}{(D-2)^2 (D-1)^2 (D+1)} + O(r^6)\,,\\
\nu &= \frac{8 \pi  \gamma^2 r^2}{D-1}-\frac{32 \pi^2 (D-3) D \gamma^4 r^4}{(D-2) (D-1)^2 (D+1)} + O(r^6)\,,
\ee\end{subequations}
which reduces to \cref{WymanIIb} when $D=4$.

%%%%%%%%%%%%%%%%%%%%%%%%%%%%
%%%%%%%%%%%%%%%%%%%%%%%%%%%%

\section*{References}
\addcontentsline{toc}{section}{References}

\bibliographystyle{iopart-num}
\bibliography{rotwym}

\end{document}